# Voltage Regulation Support Along a Distribution Line by a Virtual Power Plant Based on a Center of Mass Load Modeling

Panayiotis Moutis, *Member, IEEE*, Pavlos S. Georgilakis, *Senior Member, IEEE*, and
Nikos D. Hatziargyriou, *Fellow, IEEE*

*Abstract*—**A voltage regulation method for slow voltage variations at distribution line level is proposed, based on a view of the loads, generators and storage along a distribution line as point weights. The "centers of mass" of the absorbed and injected currents (loads & generation, respectively) are compensated by minimizing the distance between them, through proper re-dispatching of the power of the available units and interruptible loads. The technique is recursively applied to lesser parts of the distribution line to address local phenomena and is assumed to be offered as ancillary service to the system operator by a Virtual Power Plant. The favorable results of the methodology are assessed on a distribution line of the island of Rhodes (Greece) under critical loading for numerous scenarios. Unlike previous approaches, the technique focuses specifically on the restoration of bus voltages within standard limits and may reduce the activation of on-line tap changing transformers control.**

*Index Terms*—**distribution network, load modeling, smart grid, virtual power plant, voltage regulation.**

## I. INTRODUCTION

T HE past two decades, Distributed Generation (DG) and, specifically, Renewable Energy Sources (RES) have been widely and largely promoted through national and international policies and incentives [1], [2]. Despite the favorable effects of its wide penetration in modern Power Systems (PSs), DG offers little in terms of system support via ancillary services, mainly due to the reduced controllability of the units and their wide geographical and electrical dispersion.

The concerns about voltage stability and control have been considered crucial compared to other PS quality and stability issues [3]. In several countries, low voltage ride through capability of wind generators [4] and voltage profile limitations to all DG units [5] have been imposed from the early stages of their deployment. Furthermore, it was noticed that the stochastic nature of most RES-based DG units affects the Voltage Regulation (VR) offered by Transformers (T/F) equipped with On-Load Tap Changers (OLTC) [6]. Moreover and despite the provisions of connection and operation limitations, critical phenomena may still take place; i.e., considerable voltage drops, occurring locally, can lead to the disconnection of some DG units (due to the minimum operating voltage limits), which in turn causes local loads to be fed from the overlying PS, thus, further deteriorating the voltage profile of the distribution network. Although DG deployment may defer network investments (which otherwise are needed to cover the long-term load increase), the dispatching ability of DG units is a crucial issue (particularly when considering intermittent sources).

This work reviews the methods of VR at distribution level that are traditionally applied, as also the ones that are proposed in the Smart Grid (SG) related bibliography. Further, it suggests an alternative approach for slow voltage variations, which rectifies the shortcomings of previous techniques, based on the Virtual Power Plant (VPP) paradigm. The VPP concept presents several advantages, since it is abstract, relies less on the PS characteristics and control, and represents a simpler organization approach to deregulated power market models [7]. In the context of the proposed approach, a VPP offers VR as an ancillary service to the distribution system operator (DSO), through proper market mechanisms. These mechanisms are out of the scope of this study. The method views the loads along a distribution line as negative "masses" (or weights), while the DG injections as positive "masses". Hence, the DG units and the loads of the VPP, which are capable of Demand Response (DR) are controlled, so that the centers of gravity of the positive and negative "masses" are counterbalanced. Taking into account the locality of the VR problem, the method is executed along the whole distribution line, as also any lesser parts of it. The idea is based on the observation that the longest the flow of current to a load point, the greater the voltage drop; i.e., the aim is to reduce the current flowing to any load point from the overlying PS. The proposed methodology is a tool for the VPP to offer VR to the DSO, when the latter requests it.

Section II reviews thoroughly the traditional and SG related VR techniques applied to distribution networks. In Section III the proposed technique for VR support by a VPP along a distribution line is analyzed. Numerous scenarios on a realistic test system (feeder of the island of Rhodes, Greece) are used in Section IV, so as to assess the proposed control strategy. Section V discusses the applicability, prerequisites and limitations of the suggested methodology. Section VI concludes this work.

P. Moutis is with the Department of ECE, Carnegie Mellon University, USA, e-mail: pmoutis@andrew.cmu.edu.

P. S. Georgilakis and N. D. Hatziargyriou are with the School of Electrical and Computer Engineering, National Technical University of Athens (NTUA), Athens, Greece (e-mail: pgeorg@power.ece.ntua.gr; nh@power.ece.ntua.gr).

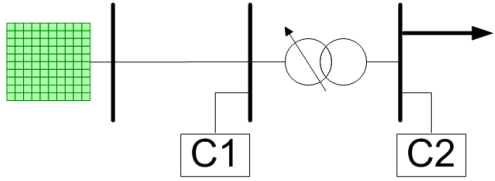

Fig. 1. Distribution line connected to the grid through an OLTC-equipped T/F and the alternative reactive compensation placements (C1 and C2).

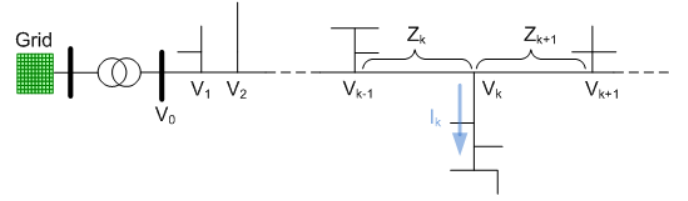

Fig. 2. Calculating injected/absorbed current of bus $k$ by neighboring buses.

## II. Review On Methods Of Voltage Regulation

### A. The Voltage Regulation Problem [8]

Any flow of current from a source to a load suffers voltage drops along the lines. In order for the voltage of all buses in distribution networks to be kept within the standardized limits [9], [10], the following principles can be identified. Firstly, the distribution feeders are properly sized to meet the current and the forecasted future loading of a certain time horizon [11]. During online operation of the distribution feeders, as shown in Fig. 1, and in order to ensure close to nominal voltages throughout the distribution network, the DSO relies (i) on the actions of the voltage controllers of the OLTC-equipped transformers and/or (ii) on the reactive power compensation to minimize voltage drops caused by the transmission and distribution system.

### B. Voltage Regulation Methods

As from II.A, most of the VR approaches are based either on OLTC control actions or on some sort of reactive power compensation.

*1) Classic VR Control:* Reactive power compensation may be realized through passive or active equipment consisting – in most cases – of capacitor banks [8], [12]. Passive topologies are activated manually, while active ones automatically adjust their reactive power injection, so as to optimize their contribution to VR.

The VR capability of OLTC-equipped T/Fs is well documented [8], [13]. However, the incremental realization and time delays of this control [14] and the various conditions under which it might cause instability [15] must be noted. Moreover, the control strategies applied translate to different effectiveness of the voltage regulation, while they may not restrain from requiring complementary support by capacitor banks [16].

All of the above methods are executed in a top-down manner by the DSO. This approach fails to serve a deregulated energy market paradigm.

*2) SG-based VR Control:* Controlling generators, storage and loads of a SG can offer VR by minimizing either PS losses (*Optimal Power Flow - OPF*) or the sum of standard deviation of bus voltages from nominal value (*voltage profile optimization*). In [14] OLTC control as complementary action on behalf of the T/F is suggested, while in some cases [17] the voltage constraints have to be relaxed until a feasible solution can be reached. Essentially, both works notice the inherent weaknesses of OPF to procure proper actions and a control strategy to regulate voltage at large. As for voltage profile optimization, it is realized by:

$$\text{minimize}: f(\tilde{x}) = \sum_{k=1}^{n}(V_k - V_{sp})^2 \qquad (1)$$

where, $V_{sp}$ is usually 1.0 per unit (p.u.), so as to minimize deviations of bus voltages from nominal. In [18] certain number of generators and loads (one per distribution line or a set behind a point of common coupling) are controlled so as to satisfy (1); the reason for this selective control technique is to avoid "hunting behavior", due to counteracting control actions [19]. Voltage constraints are not enforced in [20], but are incorporated in (1) through penalizing weights; hence, bus voltages are not guaranteed within the standardized limits.

An alternative SG methodology for VR relies on sensitivity analysis, i.e., assessment of the voltage dependent derivatives of the inverse Jacobian matrix of the Newton-Raphson power flow method (elements $N_{km}$ and $L_{km}$) as given by:

$$\begin{bmatrix} \Delta P^{(n)} \\ \Delta Q^{(n)} \end{bmatrix} = \begin{bmatrix} H & N \\ J & L \end{bmatrix} \begin{bmatrix} \Delta \theta^{(n+1)} \\ \Delta V^{(n+1)}/V^{(n)} \end{bmatrix} \qquad (2.a)$$

where,

$$\left. \begin{array}{l} H_{kk} = \dfrac{\partial \Delta P_k}{\partial \theta_k}, N_{km} = \dfrac{\partial \Delta P_k}{\partial \Delta V_m} \\[2mm] J_{km} = \dfrac{\partial \Delta Q_k}{\partial \theta_m}, L_{km} = \dfrac{\partial \Delta Q_k}{\partial V_m} \end{array} \right\} \qquad (2.b)$$

A preliminary decomposition of the studied PS in loosely interdependent areas is applied in [21], so as to following handle the maximum sensitivity of each area separately. Similarly, sensitivities of all bus voltages outside limits are each handled separately in [22]. The sum of bus voltage sensitivities is added to (1), so as the two metrics are jointly optimized in [23]. In all aforementioned applications, every time a control action is executed, the Jacobian matrix changes, and has to be recalculated and inversed anew, so as to reassess sensitivities (only [21] follows this, while [22] barely acknowledges this fact). Moreover, neither all actors in the SG can be considered controllable (so as to handle their overall effect on voltage), nor there exists a definitive rule about which sensitivities should be prioritized.

Rule-based control relies on schemes that drive the OLTC-equipped T/Fs, so as to ensure that all bus voltages are within limits. In [24] OLTC action may be complemented by DG control (that follows a sensitivity analysis as described before); however, as the authors note, the rules depend on the complexity (in terms of installed units and types of loads) of the distribution feeder. OLTC action is primarily proposed in [25] and [26], in order to enhance the penetration of DG; although the goal of these two approaches is to limit slow (long-term) voltage deviations, they fail to serve the deregulated energy model.

Relations $P \propto f$ and $Q \propto V$ are used in [27] to design a generic active/reactive power controller for DG units at distribution level; however, as previously noticed [28], the *R/X* ratio may well affect the validity of the aforementioned relations. A control strategy of coordinating all VR-capable actors is developed in [29]; OLTC-equipped T/Fs and DG units are involved in this strategy. A VR method that attempts to minimize OLTC involvement is proposed in [30], however it assumes on some considerable penetration of synchronous DG units.

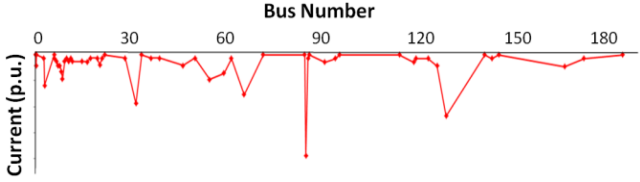

Fig. 3. Absorbed load currents along a radial distribution line of 200 buses.

### C. General Remarks on the Reviewed VR Methods

The classic approaches fail to cater for a deregulated energy market model, since they burden the DSO, in order to control the voltage, while they might have to be limited, so as not to compromise PS stability. As for the SG approaches, those based on optimization and sensitivity analysis require full access to the online PS data. OPF and voltage profile optimization fail to guarantee VR, while they might pose as over-engineering or affect severely day-ahead scheduling economics, if voltage deviations are limited to a few buses. Rule-based control views the voltage regulation from the perspective of the DSO and might require from the latter to enforce control actions to DG units.

## III. VOLTAGE REGULATION BY A VPP BASED ON MODELING OF LOADS ALONG A DISTRIBUTION LINE AS MASSES

### A. Modeling Loads as Point Masses

If smart meters cannot measure current directly, for three consecutive buses *k*-1, *k* and *k*+1 along a radial distribution line (Fig. 2), the injected/absorbed current of the middle bus *k* is given by:

$$\tilde{I}_k = \frac{\tilde{V}_k - \tilde{V}_{k-1}}{\tilde{Z}_k} - \frac{\tilde{V}_{k+1} - \tilde{V}_k}{\tilde{Z}_{k+1}} \qquad (3)$$

where voltages *V* and impedances *Z* as per Fig. 2. That said, the absorbed currents by loads along the main body of a radial distribution line (excluding lesser branches) may be represented as in Fig. 3, where the horizontal axis is the bus number of the nodes along the distribution line and the vertical axis is the per unit current absorbed. Similar depiction can be given for the injected currents from DG units. As from the above, a radial distribution network can be viewed as a body consisting of point masses of absorbed load and injected generation currents. In this scope, using as electric length of the line the accumulated impedance from the in-feeding node down to each bus, one can extract the position of the center of mass for injected and absorbed currents as the weighted average of the norm of the accumulated impedance of the buses along the line that inject and absorb current, respectively; i.e.:

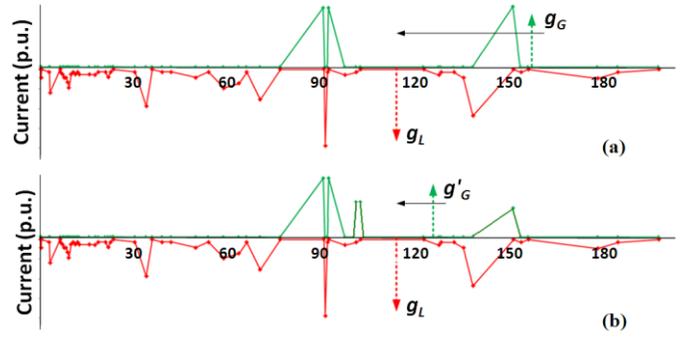

Fig. 4. Depicting the idea of compensating the centers of masses of injected/absorbed currents for the case of voltage drop (a) initial dispatching, (b) reducing DG power of bus 148 and re-dispatching to DG units on buses 108 and 109.

$$g = \frac{\sum\limits_{k=1}^{n}\left(\tilde{I}_k \sum\limits_{m=1}^{k}\left|\tilde{Z}_m\right|\right)}{\sum\limits_{k=1}^{n}\tilde{I}_k} \qquad (4)$$

In (4), the accumulated impedance $\sum\limits_{m=1}^{k}\left|\tilde{Z}_m\right|$ to each bus *k* is used as the coordinate of the current at bus *k* accounted for in the calculation.

Let the center of mass of absorbed currents be denoted as $g_L$ and of the injected ones as $g_G$. They will be arbitrarily addressed as "**symmetric**".

### B. Idea of the Proposed Method

Usually, a single operator/owner is expected to manage some of the available DG and controllable loads of a distribution line, i.e., the VPP paradigm will be considered. However, multiple VPPs may operate on a distribution line. The technique analyzed following represents a tool for each VPP to be able to offer VR when the DSO requests and may be readily applicable in a deregulated market environment.

The VR is proposed to be realized through dispatching or re-dispatching the scheduled VPP power, so that the distance between $g_L$ and $g_G$ is minimized both for the case of undervoltage and overvoltage. The idea is justified by the notion that the flow of current from the overlying grid causes losses (and thus voltage drops) along the distribution line and vice versa (for the case of voltage rise). That said and for the case of undervoltage, if the bulk of the loads (approximately represented by the corresponding center of mass $g_L$) is served "as locally as possible" by the available bulk of DG available at the distribution line (approximately represented by $g_G$), the current flows, injected in the feeder from the overlying grid, are reduced. Similar description can be given for overvoltage.

This is visualized in Fig. 4. More specifically, the injected and absorbed currents are calculated as of (3) and given as point weights in green and red, respectively, while the corresponding centers of mass are calculated as of (4) and also marked accordingly.

### C. Determining the Main Body of a Distribution Line

Generally, distribution lines are radial with branches. To account for the center of mass of the whole line, either a two-dimension calculation should be used, or each branch should be considered concentrated to its point of coupling to (what will be defined as) the main line. Commonly, the length of the branches is limited, i.e., the second approach is preferred. More specifically, Depth-First Search (DFS) [31] is firstly applied, to determine the longest electric path along the line starting from the feeder. For every bus in the path, a list of its branches (if any) is populated. The described consideration implies that the accumulated impedance $\sum_{m=1}^{k} \left| \tilde{Z}_m \right|$ to each bus $k$ of the main line is unique for that bus.

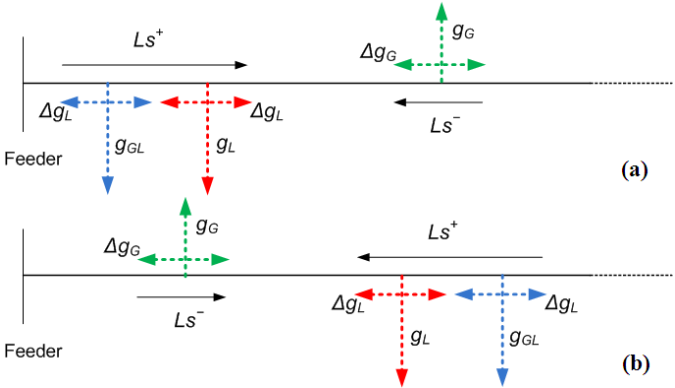

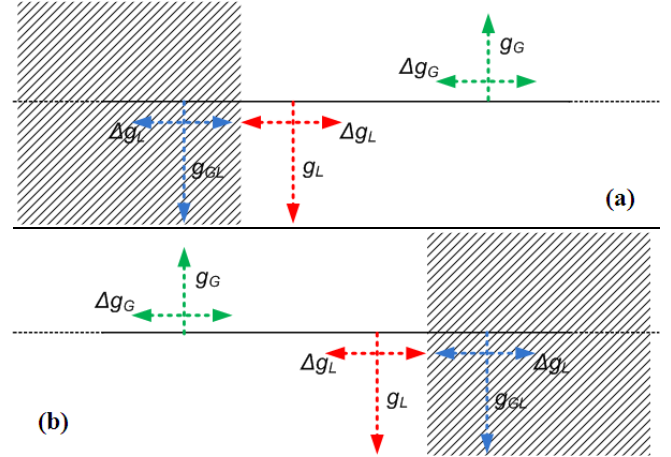

Fig. 5. Two different cases for the positions of $g_L$, $g_G$ and $g_{GL}$ considering undervoltage and the direction according to which $Ls^+$ and $Ls^-$ are populated.

### D. Analysis of the Proposed Methodology

To simplify the description of the proposed methodology the case of undervoltage will be described in the following; overvoltage may be handled accordingly.

*Calculating $g_G$, $g_L$, $g_{GL}$ and their Variance:* Assuming a smart meter infrastructure, bus currents are first gathered (direct measurements or as of (3)). Then, $g_G$, $g_L$ and $g_{GL}$ (where, $g_{GL}$ is the center of mass of the sum of the absorbed and injected currents on each bus) are calculated. The variance about the position of each center is the biased weighted estimator as:

$$\Delta g^2 = \frac{\sum_{k=1}^{n} \left( \tilde{I}_k \sum_{m=1}^{k} \left( \left| \tilde{Z}_m \right| - g \right)^2 \right)}{\sum_{k=1}^{n} \tilde{I}_k} \tag{5}$$

i.e., every center of mass will be represented as $g \pm \Delta g$.

*1) Populating (Re-)Dispatching Priority Lists:* According to the idea described in Section III.B, two lists are created, which prioritize the control actions applied to the DG units, controllable loads and storage system of the VPP. Let the first priority list be called as reduction list, $Ls^-$, and be populated by the DG units, which can curtail their output power (when re-dispatching is considered), and/or the controlled loads that may increase their consumption (storage, pumps, etc). Let the second priority list be called as increase list, $Ls+$, and be populated by DG units, which can increase their power injection, and by loads to be reduced. The following rules determine the priority of each VPP actor in the lists:

a. For undervoltage, $g_{GL}$ will be attracted closer to $g_L$ and on the opposite side of $g_G$ as per the two cases in Fig. 5.

b. Generated power at the farthest of $g_L$, but within the width of $g_G$ should be replaced by equal power closest to $g_L$

starting from the farthest within the width of $g_{GL}$.

c. Re-dispatching power from the nearest actors in between $g_G$ and $g_L$ is of the lower priority.

d. Power from units of the VPP other than the above, as also in parts of the line where $\Delta g_G$ and $\Delta g_L$ overlap are ignored.

The outlined priority rules depend on two assumptions. Firstly, the longest the distance of a VPP actor from its symmetric center of mass, the highest its priority in $Ls^-$; this holds because, according to the position of $g_G$ and $g_L$, actors of such a characteristic, affect less the position of their respective center of mass to the opposite direction of the one preferred and more their symmetric one towards the preferred direction. Secondly, $g_{GL}$ and $g_L$ (plus widths) define the window within which DG power or load shedding should be dispatched. This assumption is valid since $g_{GL}$ is as farther from $g_L$, as $g_L$ is less compensated by injected (generated) power along the distribution line. The analysis presented in this paragraph may be followed also through its graphic representation in Fig. 5.

The availability of power from stochastic RES-based DG units will be assessed according to some forecast or estimation [32].

*2) Power (Re-)Dispatching Step:* The **step** of power is the amount to be de-loaded from the first VPP actor(s) in $Ls^-$ and to be increased in the first VPP actor(s) in $Ls^+$, provided this action improves the minimum voltage in the PS or the voltage profile, accordingly. Otherwise, the technique will move to lesser parts of the distribution line (see Section III.D.4). Every control action affects the position of $g_G$, $g_L$ and $g_{GL}$; i.e., the (re-)dispatching step should be small enough so as to avoid retrogression among the position of the centers of masses (similar to "hunting behavior" [19]). Nevertheless, this may be handled by readjusting the step according to the effect of the compensation; e.g., binary-type adjustment, i.e., if compensation is small, then double the step, etc. Moreover, due to the negligible method complexity the step can be kept small.

In case either $Ls^-$ or $Ls^+$ are empty or the available power represented in either of them is less than **step**, the amount of power required to be (re-)dispatched is noted as "**rest**" and added to the next iteration of the methodology.

Although re-dispatching of the reactive power of the DG

units and loads could be an option, this work is based on present energy market framework and preferred policies and grid codes [33]–[35] that tend to restrict reactive injections from distributed energy resources. For the case of [36], which assumes on the capability of DG units to inject reactive power for the purpose of VR support, the allowed contribution is limited, while its effect may be negligible (especially for low-voltage networks) according to the observations in [28]. Such an option, however, could be readily incorporated to the technique through proper consideration of the R/X ratio of the distribution line [28] and/or of the power factor of the injected/absorbed currents. Essentially, the proposed method represents the worst case scenario for a VPP offering VR as an ancillary service, based only on the re-dispatching of its active power.

*3) Applying Proposed VR to Lesser Parts of the Distribution Line:* If $g_G$ and $g_L$ approach sufficiently (to the effect of rule (d) of Section III.D.2) or $g_{GL}$ is found between $g_G$ and $g_L$ or the voltage profile is not improved after the last (re-)dispatching, then the main body of the distribution line cannot be compensated further and the proposed VR will be applied to lesser parts of the distribution line (i.e., part of it will be ignored), defined according to the following rules:

a.  If $g_G > g_L$, adjust between $g_L - \Delta g_L$ and termination of the assumed lesser part of the distribution line as per Fig. 6(a).
b.  If $g_G < g_L$, adjust between beginning of the assumed lesser part of the distribution line and $g_G + \Delta g_G$ as per Fig. 6(b).

According to the described methodology for adjusting the assumed lesser parts of the distribution line, the technique is applied recursively so as to account for total VR. If the length of the assumed lesser part of the distribution line falls below some threshold, then the technique is executed a second time. The proposed control stops at any point all voltages or some mean of them are within standardized/preferred limits. This implies that although the method is focused on the main body of the distribution line, all bus voltages are monitored to account for the effectiveness of the offered VR.

### E. General Remarks on the Proposed Methodology

The method is not optimal and is affected by the initial scheduling of the VPP, as also the loading of the distribution line. It also has limited capability depending on the size of the VPP compared to the total load of the distribution line. The above explains the conditions under which the technique may fail to offer sufficient VR. However, since it is offered as an ancillary service to the DSO, the latter may decide to address additional VPP(s) operating on the same distribution line, before resorting to traditional methods, such as activation of the OLTC control. Let it be noted though that the system operator may decide to prioritize the use of the OLTC over the proposed technique according to proper assessment of economics or technical reasons (availability or cases like the ones described in [6]).

The proposed technique does not require additional reactive power injection (which burdens the lines with reactive current), does not involve OLTC action of the T/F, may be offered as ancillary service on behalf of the VPP to the DSO (based on limited information), improves steady state stability although it restrains from *a priori* load shedding, does not overachieve VR through voltage profile optimization or minimization of PS losses, while it may (as shown in Section IV) re-dispatch the scheduled VPP power without resorting to additional DG injections or load shedding, thus minimizing the effect of VR on the economical operation. Obviously, critical availability on behalf of stochastic RES-based DG units could be restricted from being curtailed by the proposed technique or shifted through proper control of the VPP storage systems. The challenge of fast and accurate measurements on behalf of the smart meter infrastructure has to be acknowledged, however both standards [37] and current technologies [38] can readily handle the above and facilitate the proposed method. The complete flow of the proposed method is given in Fig. 7.

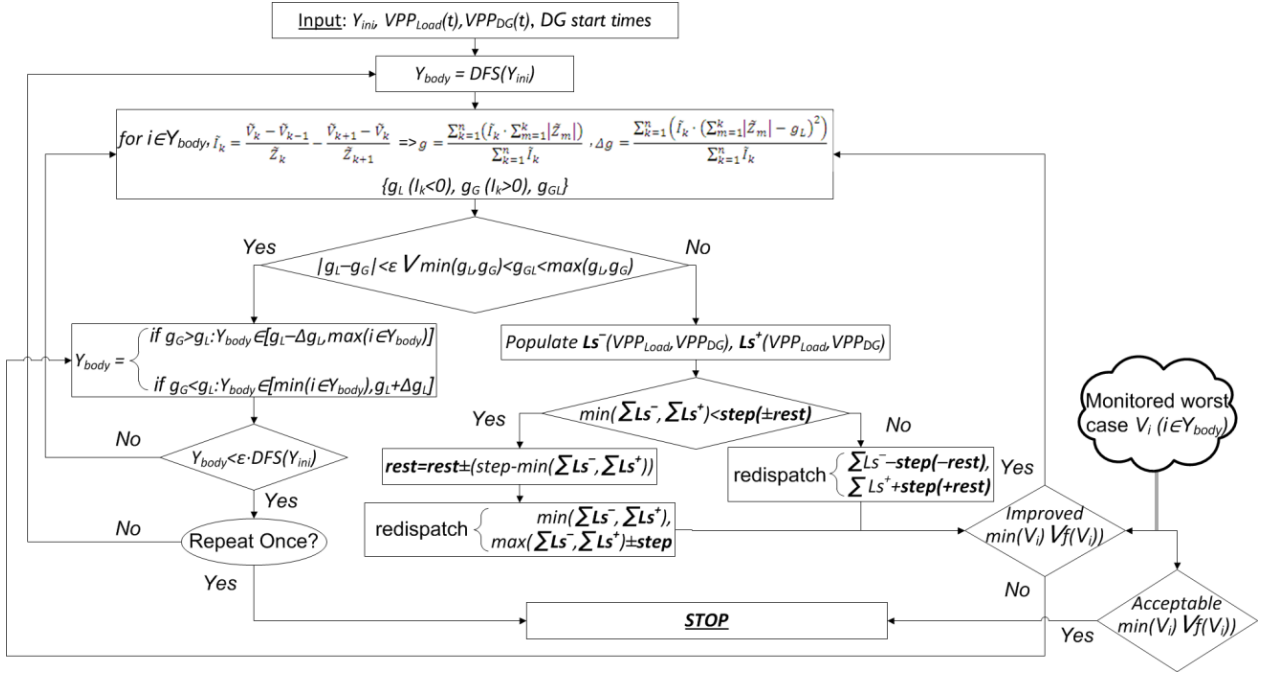

Fig. 7. Flow of the algorithm of the proposed VR support method by a VPP for distribution lines. ($Y$: admittance matrix of the distribution line, $VPP_{load}(t)$ and $VPP_{DG}(t)$: initial scheduling of the VPP actors, $Y_{body}$: assumed distribution line to which re-dispatching is applied, rest as described in Section III).

**TABLE I**
VPP Considered Operating on the
R-26 Distribution Feeder of the Island of Rhodes (Greece)

| Bus | Unit type | $P_n$ (kW) | Bus | Unit type | $P_n$ (kW) |
|-----|-----------|-----------|-----|-----------|-----------|
| 213 | Wind Park | 200 | 204 | IL | 120 |
| 224 | Photovoltaic-Hydro | 150 | 358 | IL | 120 |
| 316 | Wind Park | 220 | 369 | IL | 120 |
| 324 | Wind Park | 150 | 291 | IL | 120 |
| 324 | Biomass | 250 | 220 | IL | 120 |
| 324 | LPG Gen. | 150 | 222 | IL | 120 |
| 380 | Photovoltaic-Hydro | 300 | 237 | IL | 80 |
| 414 | Geothermal | 200 | 239 | IL | 80 |
| 376 | LPG Gen. | 150 | 249 | IL | 80 |
| 381 | Diesel | 150 | 259 | IL | 80 |
| 384 | Diesel | 200 | 328 | IL | 80 |
| 227 | IL | 150 | | | |

## IV. VR Support by a VPP on the Rhodes Island of Greece Test System

Feeder R-26 (Gennadi) of the PS of the island of Rhodes in Greece is considered as the test system for the proposed methodology. It is mostly radial and consists of 119 load buses representing a total peak load of 8.1MW. The main part of the distribution line is of CU-95 type, while the branches are of ACSR-35, ACSR-16, AAAC-35, CU-35 and CU-16 types. The nodes are numbered between 200-433 (excess buses are branch nodes and poles). Further details of the test system are given in the Appendix.

The VPP described in Table I is considered to be operating on the test system. Its total peak power is 3.4MW, 2.1MW of which is DG and 1.3MW of which is interruptible load (IL).

For the assumed PS and the considered VPP a large number of initial loading scenarios were randomly generated and then the proposed technique was applied and assessed.

**TABLE II**
Statistics of Successful VR Support by VPP on the Island of Rhodes
for 300KW Re-Dispatching Step

| | Loading (kW) | | | | Number of iterations |
|---|---|---|---|---|---|
| | R-26 initially | VPP initially | VPP change | IL shed | |
| **Average** | 4021 | 1056 | 343 | 333 | 3.59 |
| **Std. Dev.** | 101 | 190 | 278 | 169 | 1.25 |
| **Maximum** | 4341 | 1595 | 1052 | 800 | 7.00 |
| **Minimum** | 3763 | 609 | 0 (18%) | 42 | 2.00 |

**TABLE III**
Statistics of Failed VR Support by VPP on the Island of Rhodes for
300KW Re-Dispatching Step

| | Loading (kW) | | | | Number of iterations |
|---|---|---|---|---|---|
| | R-26 initially | VPP initially | VPP change | IL shed | |
| **Average** | 4055 | 1031 | 113 | 96 | 3.34 |
| **Std. Dev.** | 102 | 173 | 225 | 161 | 0.98 |
| **Maximum** | 4390 | 1613 | 1086 | 693 | 7.00 |
| **Minimum** | 3744 | 552 | 0 | 0 | 1.00 |

**TABLE IV**
Statistics of Successful VR Support by VPP on the Island of Rhodes
for 300KW Dispatching Step

| | Loading (kW) | | | Number of iterations |
|---|---|---|---|---|
| | R-26 initially | VPP change | IL shed | |
| **Average** | 3972 | 1191 | 444 | 7.65 |
| **Std. Dev.** | 126 | 262 | 109 | 1.61 |
| **Maximum** | 4275 | 1916 | 896 | 12.00 |
| **Minimum** | 3563 | 507 | 216 | 3.00 |

**TABLE V**
Statistics of Failed VR Support by VPP on the Island of Rhodes for
300KW Dispatching Step

| | Loading (kW) | | | Number of iterations |
|---|---|---|---|---|
| | R-26 initially | VPP change | IL shed | |
| **Average** | 4024 | 1008 | 376 | 6.65 |
| **Std. Dev.** | 115 | 263 | 100 | 2.07 |
| **Maximum** | 4390 | 1838 | 718 | 12.00 |
| **Minimum** | 3702 | 517 | 201 | 4.00 |

More specifically, the level of loading that would cause voltage drop to one or more buses below 0.9 p.u. was approximated (let it be called **critical loading**).

Following, based on Monte Carlo simulations, a large number of loading scenarios was generated around the critical loading. As for the VPP, it would be considered either not injecting any power to the PS (VR through VPP dispatching) or scheduled at a random initial loading (VR through VPP re-dispatching). The method is applied for both initial states of the VPP.

According to what was discussed in Section III.D.3, the (re-)dispatching step will be the parameter to be concerned and discussed. Although artificial intelligence methods, such as Artificial Neural Networks, Decision Trees and Fuzzy Logic could be employed to adjust the step of the technique, a constant re-dispatching step is selected. Any (re-)dispatching step below 350kW proved to offer negligible VR support; that is why the steps 300kW, 400kW and 500kW are presented.

### A. Proposed Method of VR Support by VPP on the Island of Rhodes for a 300kW (re-)dispatching step

For the case of re-dispatching a random initial scheduling of the VPP so as to support VR in the PS, it was found that the technique was successful in 27.66% of the various PS loading scenarios and failed in the rest. For the successful VR on behalf of the method, the statistics in Table II were gathered and are indicatively outlined in detail for the reader's ease.

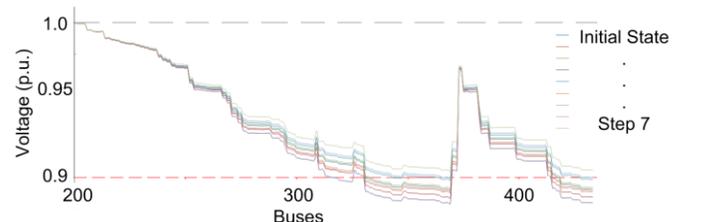

Fig. 8. Voltage profile improvement per each step of re-dispatching according to the proposed methodology for a 300kW step.

TABLE VI
STATISTICS OF SUCCESSFUL VR SUPPORT BY VPP ON THE ISLAND OF RHODES FOR 400KW RE-DISPATCHING STEP

| | Loading (kW) | | | | Number of iterations |
| | R-26 initially | VPP initially | VPP change | IL shed | |
|---|---|---|---|---|---|
| **Average** | 4023 | 1024 | 410 | 328 | 3.35 |
| **Std. Dev.** | 96 | 194 | 309 | 189 | 1.24 |
| **Maximum** | 4304 | 1719 | 1598 | 836 | 9.00 |
| **Minimum** | 3720 | 462 | 0 (17%) | 0 | 2.00 |

From all Monte Carlo generated loading scenarios, the average loading of the feeder was 4021kW, with a standard deviation of 101kW, maximum loading of 4341kW and minimum of 3763kW. The scheduled VPP power before employing the proposed VR method was on average 1056kW, with a standard deviation of 190kW and maximum and minimum power of 1595kW and 609kW, respectively. Executing the proposed method led to an average total additional VPP power injection of 343±278kW, which in the worst case required 1052kW of extra injection, while in 18% of all successful re-dispatching control actions the total VPP active power change was zero. The interruptible load employed by the method averaged at 333±169kW, with a maximum demand response of 800kW and minimum of 42kW. The technique required an average of 3.59 iterations (re-dispatching steps) ±1.25 iterations to succeed in offering VR within acceptable limits, with a maximum of seven and a minimum of two iterations (steps). Fig. 8 shows the step by step improvement of the voltage profile of the proposed methodology for the 300kW step. Table III presents the statistics of failed VR support by the technique.

For the case of dispatching the VPP so as to support VR, although not scheduled in the first place, the methodology was successful in 39.2% of the initial loading scenarios of the PS. For the successful and failed instances, the statistics are presented in Tables IV and V, respectively.

The following conclusions are drawn from Tables II to V:

a. Table II shows that the average number of re-dispatching iterations is 3, i.e., 900kW of VPP power are re-dispatched.

TABLE VII
STATISTICS OF FAILED VR SUPPORT BY VPP ON THE ISLAND OF RHODES FOR 400KW RE-DISPATCHING STEP

| | Loading (kW) | | | | Number of iterations |
| | R-26 initially | VPP initially | VPP change | IL shed | |
|---|---|---|---|---|---|
| **Average** | 4073 | 1044 | 264 | 173 | 3.38 |
| **Std. Dev.** | 101 | 179 | 353 | 205 | 1.03 |
| **Maximum** | 4422 | 1452 | 1234 | 764 | 7.00 |
| **Minimum** | 3827 | 571 | 0 | 0 | 1.00 |

TABLE VIII
STATISTICS OF SUCCESSFUL VR SUPPORT BY VPP ON THE ISLAND OF RHODES FOR 400KW DISPATCHING STEP

| | Loading (kW) | | | Number of iterations |
| | R-26 initially | VPP change | IL shed | |
|---|---|---|---|---|
| **Average** | 4005 | 1248 | 456 | 6.00 |
| **Std. Dev.** | 115 | 261 | 107 | 1.14 |
| **Maximum** | 4349 | 2176 | 886 | 12.00 |
| **Minimum** | 3669 | 400 | 227 | 2.00 |

TABLE IX
STATISTICS OF FAILED VR SUPPORT BY VPP ON THE ISLAND OF RHODES FOR 400KW DISPATCHING STEP

| | Loading (kW) | | | Number of iterations |
| | R-26 initially | VPP change | IL shed | |
|---|---|---|---|---|
| **Average** | 4003 | 1090 | 328 | 5.02 |
| **Std. Dev.** | 154 | 362 | 128 | 1.77 |
| **Maximum** | 4422 | 2188 | 702 | 11.00 |
| **Minimum** | 3699 | 600 | 173 | 3.00 |

TABLE X
STATISTICS OF SUCCESSFUL VR SUPPORT BY VPP ON THE ISLAND OF RHODES FOR 500KW RE-DISPATCHING STEP

| | Loading (kW) | | | | Number of iterations |
| | R-26 initially | VPP initially | VPP change | IL shed | |
|---|---|---|---|---|---|
| **Average** | 4025 | 1025 | 382 | 272 | 2.97 |
| **Std. Dev.** | 103 | 177 | 306 | 197 | 1.05 |
| **Maximum** | 4352 | 1534 | 1733 | 938 | 8.00 |
| **Minimum** | 3737 | 443 | 0 (10%) | 42 | 2.00 |

Nevertheless, the average change in VPP power is much less and the contribution of IL is around 300kW. This means that VR is achieved with internal re-dispatching of the generation of the VPP only.

b. Comparing Tables II and III, as also Tables IV and V, one can notice that the success or failure of the technique is independent of the amount of initial loading of the distribution line or the VPP, hence the method depends on the initial *allocation* of the loads and/or VPP scheduling.

c. The average power of dispatching the VPP in order to offer VR support (Table IV) is less compared to re-dispatching it (equal to the average VPP power change plus its initial loading as of Table II).

Due to space limitations, only the tables of the statistics for each of the rest of the re-dispatching steps will be presented in IV.B to IV.D. The comments are similar and general remarks will, following, summarize the findings gathered.

### B. Proposed Method of VR Support by VPP on the Island of Rhodes for a 400kW (re-)dispatching step

The re-dispatching of the initial VPP scheduling yields 59.28% success rate for a step of 400kW. Employing the VPP

TABLE XI
STATISTICS OF FAILED VR SUPPORT BY VPP ON THE ISLAND OF RHODES FOR 500KW RE-DISPATCHING STEP

| | Loading (kW) | | | | Number of iterations |
| | R-26 initially | VPP initially | VPP change | IL shed | |
|---|---|---|---|---|---|
| **Average** | 4113 | 1024 | 473 | 273 | 3.35 |
| **Std. Dev.** | 104 | 193 | 375 | 199 | 1.19 |
| **Maximum** | 4362 | 1395 | 1388 | 736 | 8.00 |
| **Minimum** | 3837 | 468 | 0 | 0 | 1.00 |

TABLE XII
STATISTICS OF SUCCESSFUL VR SUPPORT BY VPP ON THE ISLAND OF RHODES FOR 500KW DISPATCHING STEP

| | Loading (kW) | | | Number of iterations |
| | R-26 initially | VPP change | IL shed | |
|---|---|---|---|---|
| **Average** | 4002 | 1296 | 473 | 5.05 |
| **Std. Dev.** | 120 | 240 | 96 | 0.83 |
| **Maximum** | 4362 | 2186 | 857 | 9.00 |
| **Minimum** | 3678 | 500 | 216 | 2.00 |

### TABLE XIII
#### Statistics of Failed VR Support by VPP on the Island of Rhodes for 500KW Dispatching Step

| | Loading (kW) | | | Number of iterations |
|---|---|---|---|---|
| | R-26 initially | VPP change | IL shed | |
| **Average** | 4029 | 1301 | 394 | 4.78 |
| **Std. Dev.** | 142 | 412 | 134 | 2.20 |
| **Maximum** | 4338 | 2360 | 661 | 9.00 |
| **Minimum** | 3773 | 1000 | 207 | 3.00 |

### TABLE XIV
#### Total Average Power of the VPP and Contribution of the IL With Regards to the Re-Dispatching Step of the Methodology

| | Re-Dispatching Step (kW) | | |
|---|---|---|---|
| | 300 | 400 | 500 |
| **Total VPP power (kW)** | 1056(±190)+ +343(±278) | 1024(±194)+ +410(±309) | 1025(±177)+ +382(±306) |
| **IL shed (kW)** | 333(±169) | 328(±189) | 272(±197) |

just for VR support with the same amount of power dispatching step is successful in 88.73% of all initial loadings. The statistics for the (re-)dispatching step of 400kW are given in Tables VI to IX.

### C. Proposed Method of VR Support by VPP on the Island of Rhodes for a 500kW (re-)dispatching step

For the case of re-dispatching the initial VPP scheduling, the technique was successful in 82.69% of the various PS loading scenarios and failed in the rest, while for the case of dispatching the VPP just for VR support the success rate was 96.09%. The statistics for a (re-)dispatching step of 500kW are outlined in the following Tables X to XIII.

It can be noticed that there is a considerable decrease in the cases that achieve VR with zero change in the output of the VPP power. For a re-dispatching step of 500kW, only 10% of the cases do not require additional power on behalf of the VPP, while for steps of 300kW and 400kW that amount was almost double (18%).

### D. Comments on the Effect of the Re-Dispatching Step to the Efficiency of the Proposed VR Support Methodology by a VPP

Assessing the results of Sections IV.A to IV.C, the following comments can be made:

a. The total power required on behalf of the VPP both for the cases of re-dispatching and dispatching its availability in order to offer VR is independent of the (re-)dispatching step. The same can be noticed for the amount of IL shed.

### TABLE XV
#### Total Average Power of the VPP and Contribution of the IL With Regards to the Dispatching Step of the Methodology

| | Dispatching Step (kW) | | |
|---|---|---|---|
| | 300 | 400 | 500 |
| **Total VPP power (kW)** | 1191(±262) | 1248(±261) | 1296(±240) |
| **IL shed (kW)** | 444(±109) | 456(±107) | 473(±96) |

### TABLE XVI
#### Success Rates of the Technique With Regards to the Re-Dispatching Step

| | Re-Dispatching Step (kW) | | |
|---|---|---|---|
| | 300 | 400 | 500 |
| **Re-dispatching Success (%)** | 27.66 | 59.28 | 82.69 |
| **Dispatching Success (%)** | 39.20 | 88.73 | 96.09 |

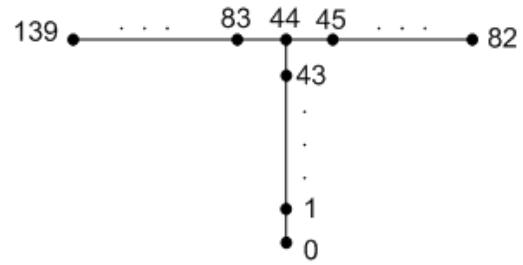

Fig. 9. Single line representation of one of the island of Ikaria distribution lines, depicting a network, which may not be considered mostly radial.

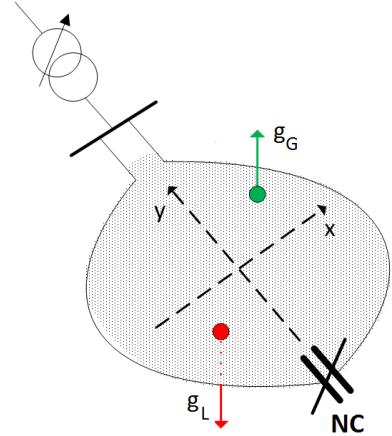

Fig. 10. Example of two distribution lines connected in a loop through a normally closed switch and the xy coordinate system over which the location of the centers of mass of load and generation are determined to employ the proposed VR technique.

Tables XIV and XV support this observation.

b. As the (re-)dispatching step increases, the success rate of the method increases. This is summarized in Table XVI.

c. As the (re-)dispatching step increases, the method requires less iterations to termination.

d. Taking into account the above comments (b), (c) as also the observation that the *allocation* of the initial loading (either of the feeder or of the VPP) affects the proposed method, the success rate of the greater (re-)dispatching steps can be explained as following. It is the center of mass of the entire distribution line (or, at least, of its greater parts as these are assumed after applying the first adjustments described in Section III.D.4) that mostly needs to be accounted for, in order for the VPP to successfully offer VR support. That said, compensating the centers of mass of lesser parts of the distribution line has minor effect (i.e., the later adjustments according to Section III.D. 4).

## V. Limitations to the Applicability of the Method

First issue to be observed is that the proposed methodology requires access to online metering data while the VPP owner/operator may be reluctant to disclose the operational characteristics of its actors to a DSO or network operator. However, this may be rectified by integrating the technique in a web-service that ensures privacy for both (or more) parties involved. It should be also noted that unlike OPF and sensitivity-based techniques, the proposed method restrains from requiring knowledge of all loads and DG units in the distribution line (excluding those of the VPP).

TABLE XVII
RATED POWER OF MV-LV DISTRIBUTION TRANSFORMERS CONNECTED
ALONG R-260 FEEDER OF THE ISLAND OF RHODES POWER SYSTEM

| Node | $P_{T/F}$ (kW) | Node | $P_{T/F}$ (kW) | Node | $P_{T/F}$ (kW) | Node | $P_{T/F}$ (kW) |
|------|------|------|------|------|------|------|------|
| 204 | 160 | 248 | 50 | 288 | 100 | 334 | 50 |
| 207 | 50 | 249 | 100 | 291 | 160 | 336 | 100 |
| 211 | 100 | 251 | 50 | 292 | 100 | 337 | 50 |
| 212 | 100 | 253 | 50 | 294 | 50 | 342 | 50 |
| 213 | 250 | 256 | 100 | 297 | 25 | 344 | 160 |
| 216 | 50 | 258 | 100 | 299 | 50 | 346 | 50 |
| 218 | 100 | 259 | 100 | 301 | 50 | 347 | 100 |
| 220 | 160 | 261 | 100 | 303 | 100 | 349 | 50 |
| 222 | 160 | 263 | 100 | 305 | 25 | 351 | 50 |
| 224 | 250 | 265 | 100 | 307 | 50 | 353 | 50 |
| 226 | 100 | 266 | 100 | 308 | 50 | 355 | 50 |
| 227 | 250 | 268 | 50 | 314 | 50 | 356 | 160 |
| 229 | 100 | 270 | 50 | 316 | 400 | 358 | 160 |
| 231 | 50 | 272 | 50 | 318 | 50 | 364 | 50 |
| 233 | 100 | 273 | 100 | 320 | 50 | 366 | 50 |
| 235 | 50 | 275 | 50 | 322 | 50 | 368 | 50 |
| 237 | 100 | 277 | 50 | 324 | 734 | 369 | 160 |
| 239 | 100 | 279 | 50 | 325 | 50 | 371 | 50 |
| 241 | 100 | 281 | 100 | 328 | 100 | 389 | 100 |
| 243 | 50 | 283 | 50 | 330 | 50 | 411 | 100 |
| 245 | 50 | 285 | 100 | 332 | 50 | 418 | 100 |
| 204 | 160 | 248 | 50 | 288 | 100 | 334 | 50 |
| 207 | 50 | 249 | 100 | 291 | 160 | 336 | 100 |
| 211 | 100 | 251 | 50 | 292 | 100 | 337 | 50 |
| 212 | 100 | 253 | 50 | 294 | 50 | 342 | 50 |
| 213 | 250 | 256 | 100 | 297 | 25 | 344 | 160 |
| 216 | 50 | 258 | 100 | 299 | 50 | 346 | 50 |
| 218 | 100 | 259 | 100 | 301 | 50 | 347 | 100 |

Secondly, the tests presented here concern a mostly radial distribution feeder with much lesser (in length and loading) branches. In [7] the feeder of Ikaria, at half its length, splits in two major branches of equal length and loading (as per Fig. 9). Although such cases are rare, the method is still applicable by expanding to a second dimension and considering $\Delta g$ as the radius of a circle centered at the corresponding $g$. Further to this, a more interesting (although particularly rare) topology is that of loop distribution lines. The main body of the line, as described in Section III.C, could then be viewed as a circle the center of which would be the origin of the coordinate system. The centers of mass of the injected/absorbed currents could then be viewed as shown in Fig. 10 and the technique could be similarly realized. For any additional distribution lines loop to the above topology, an additional dimension should be assumed and the technique could be expanded accordingly.

Although the market mechanism for incentivizing and operating the proposed VR ancillary service offered by VPPs to DSOs is out of the scope of this work, it has to be noticed that in every case, the VPP will be injecting at least the same total active power, which was injected before its re-dispatching, i.e., there are no energy revenue losses for the VPP. On the contrary, the VPP will have additional revenue, provided a proper pricing for the service is designed.

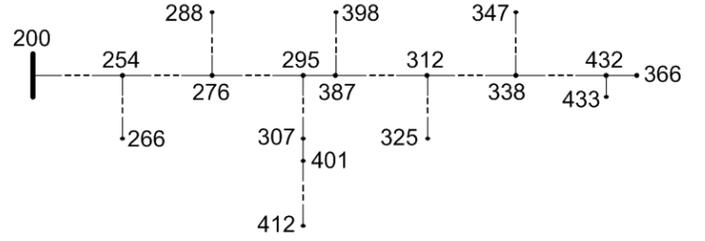

Fig. 11. Single line equivalent of the R-26 (Gennadi) feeder of the PS of the island of Rhodes (Greece).

## VI. CONCLUSION

A novel technique for the voltage support along a distribution line has been proposed. The technique assumes on the VPP paradigm and is based on the modeling of demand and generation as negative and positive weights (absorbed and injected currents), respectively. The idea of the proposed method is that, for the case of undervoltage, the closer the generation to the load, then the requirement for current flow from the overlying grid is reduced. This is expressed by the objective of the methodology to counter-balance the centers of masses of the aforementioned weights by (re-)dispatching the VPP load and generation. The suggested control strategy is applied to the full extent of the distribution line, as also to its lesser parts, in order to address the locality of the voltage phenomena. Compared to previous realizations of voltage regulation, the proposed technique is favorable at multiple economical and technical levels. The results show that the parameter of the (re-)dispatching step has to be tuned accordingly for improved performance, although the total (re-)dispatched VPP power is not affected by it. Some policy and disclosure issues might arise, but there are means for those to be rectified. Including reactive power control in the proposed method could be examined in view of changes in grid codes.

## APPENDIX

The single line equivalent of the R-26 (Gennadi) feeder of the PS of the island of Rhodes in Greece is presented in Fig. 11. The branches longer than 3 nodes along the feeder are also indicatively marked. The rated power of the medium voltage (MV) to low voltage (LV) distribution transformers is given in Table XVII.

## VII. REFERENCES

[1] V. Madani *et al.*, "Distribution Automation Strategies Challenges and Opportunities in a Changing Landscape," *IEEE Trans. Smart Grid*, vol. 6, no. 4, pp. 2157–2165, July 2015.

[2] M. Džamarija and A. Keane, "Autonomous Curtailment Control in Distributed Generation Planning," *IEEE Trans. Smart Grid*, vol. 7, no. 3, pp. 1337–1345, May 2009.

[3] *IEEE Recommended Practice for Monitoring Electric Power Quality*, IEEE Std. 1159-2009, Jun. 2009.

[4] P. Mitra, G. K. Venayagamoorthy, and K. A. Corzine, "SmartPark as a Virtual STATCOM," *IEEE Trans. Smart Grid*, vol. 2, no. 3, pp. 445–455, Sept. 2011.

[5] M. Tsili and S. Papathanassiou, "A review of grid code technical requirements for wind farms," *IET Renew. Power Gener.*, vol. 3, no. 3, pp. 308–332, Sept. 2009.

[6] R. R. Londero, C. d. M. Affonso and J. P. A. Vieira, "Long-Term Voltage Stability Analysis of Variable Speed Wind Generators," *IEEE Trans. Power Syst.*, vol. 30, no. 1, pp. 439–447, Jan. 2015.

[7] P. Moutis and N. D. Hatziargyriou, "Decision Trees-Aided Active Power Reduction of a Virtual Power Plant for Power System Over-Frequency Mitigation," *IEEE Trans. Ind. Informat.*, vol. 11, no. 1, pp. 251–261, Feb. 2015.

[8] P. Kundur, *Power System Stability and Control*. New York: McGraw-Hill, 1993.

[9] *IEC Standard Voltages*, Eur. Std. IEC 60038:2009, Jun. 2009.

[10] *Electric Power Systems and Equipment-Voltage Ratings (60 Hertz)*, American National Standards Institute (ANSI) C84.1-1995 (R2005).

[11] E. Lakervi and E. J. Holmes, *Electricity Distribution Network Design*. London, U.K.: Inst. Eng. Technol., 2003.

[12] T. J. E. Miller, *Reactive Power Control in Electric Power Systems*. New York: Wiley, 1982.

[13] T. Van Cutsem, C. Vournas, *Voltage Stability of Electric Power Systems*. Kluwer, 1998.

[14] Z. Ziadi, S. Taira, M. Oshiro and T. Funabashi, "Optimal Power Scheduling for Smart Grids Considering Controllable Loads and High Penetration of Photovoltaic Generation," *IEEE Trans. Smart Grid*, vol. 5, no. 5, pp. 2350–2359, Sept. 2014.

[15] T. Van Cutsem, "Voltage instability: phenomena, countermeasures, and analysis methods," *Proc. IEEE*, vol. 88, no. 2, pp. 208–227, Feb. 2000.

[16] C. Long and L. F. Ochoa, "Voltage Control of PV-Rich LV Networks: OLTC-Fitted Transformer and Capacitor Banks," *IEEE Trans. Power Syst.*, article in press.

[17] Q. Guo, H. Sun, M. Zhang, J. Tong, B. Zhang and B. Wang, "Optimal Voltage Control of PJM Smart Transmission Grid: Study, Implementation, and Evaluation," *IEEE Trans. Smart Grid*, vol. 4, no. 3, pp. 1665–1674, Sept. 2013.

[18] A. R. Di Fazio, G. Fusco and M. Russo, "Decentralized Control of Distributed Generation for Voltage Profile Optimization in Smart Feeders," *IEEE Trans. Smart Grid*, vol. 4, no. 3, pp. 1586–1596, Sept. 2013.

[19] P. N. Vovos, A. E. Kiprakis, A. R. Wallace and G. P. Harrison, "Centralized and Distributed Voltage Control: Impact on Distributed Generation Penetration," *IEEE Trans. Power Syst.*, vol. 22, no. 1, pp. 476–483, Feb. 2007.

[20] A. A. Aquino-Lugo, R. Klump and T. J. Overbye, "A Control Framework for the Smart Grid for Voltage Support Using Agent-Based Technologies," *IEEE Trans. Smart Grid*, vol. 2, no. 1, pp. 173–180, Mar. 2011.

[21] L. Yu, D. Czarkowski and F. de Leon, "Optimal Distributed Voltage Regulation for Secondary Networks With DGs," *IEEE Trans. Smart Grid*, vol. 3, no. 2, pp. 959–967, June 2012.

[22] M. Brenna *et al.*, "Automatic Distributed Voltage Control Algorithm in Smart Grids Applications," *IEEE Trans. Smart Grid*, vol. 4, no. 2, pp. 877–885, June 2013.

[23] K. Christakou, D. C. Tomozei, J. Y. Le Boudec and M. Paolone, "GECN: Primary Voltage Control for Active Distribution Networks via Real-Time Demand-Response," *IEEE Trans. Smart Grid*, vol. 5, no. 2, pp. 622–631, March 2014.

[24] A. Kulmala, S. Repo and P. Järventausta, "Coordinated Voltage Control in Distribution Networks Including Several Distributed Energy Resources," *IEEE Trans. Smart Grid*, vol. 5, no. 4, pp. 2010–2020, July 2014.

[25] C. M. Hird, H. Leite, N. Jenkins and H. Li, "Network voltage controller for distributed generation," *IEE Proc. Gener. Transm. Distrib.*, vol. 151, no. 2, pp. 150–156, Mar. 2004.

[26] T. Pfajfar, I. Papic, B. Bletterie and H. Brunner, "Improving power quality with coordinated voltage control in networks with dispersed generation," in *Int. Conf. Electrical Power Quality and Utilisation*, Barcelona, 2007, pp. 1–6.

[27] H. Bevrani and S. Shokoohi, "An Intelligent Droop Control for Simultaneous Voltage and Frequency Regulation in Islanded Microgrids," *IEEE Trans. Smart Grid*, vol. 4, no. 3, pp. 1505–1513, Sept. 2013.

[28] A. Engler and N. Soultanis, "Droop control in LV-grids," in *Int. Conf. Future Power Syst.*, Amsterdam, 2005, pp. 1–6.

[29] M. S. El Moursi, H. H. Zeineldin, J. L. Kirtley and K. Alobeidli, "A Dynamic Master/Slave Reactive Power-Management Scheme for Smart Grids With Distributed Generation," *IEEE Trans. Power Del.*, vol. 29, no. 3, pp. 1157–1167, June 2014.

[30] Y. J. Kim, J. L. Kirtley and L. K. Norford, "Reactive Power Ancillary Service of Synchronous DGs in Coordination With Voltage Control Devices," *IEEE Trans. Smart Grid*, article in press.

[31] R. Tarjan, "Depth-first search and linear graph algorithms," *SIAM J. Computing*, vol. 1, no. 2, pp. 146–160, 1972.

[32] G. Sideratos and N. D. Hatziargyriou, "Probabilistic Wind Power Forecasting Using Radial Basis Function Neural Networks," in *IEEE Transactions on Power Systems*, vol. 27, no. 4, pp. 1788-1796, Nov. 2012.

[33] *Power Generation Systems Connected to the Low-Voltage Distribution Network—Technical Minimum Requirements for the Connection to and Parallel Operation With Low-Voltage Distribution Networks*, document VDE-AR-N 4105:2011-08, VDE Verband der Elektrotechnik Elektronik Informationstechnik e.V., 2010.

[34] *Characteristic of the Utility Interface for Photovoltaic (PV) Systems*, IEC Std. 61727, 2002.

[35] *IEEE Application Guide for IEEE Std 1547(TM), IEEE Standard for Interconnecting Distributed Resources with Electric Power Systems*, pp. 1–217, 2009.

[36] Bartels, Wolfgang, et al. "Generating plants connected to the medium-voltage network." *Technical Guideline of BDEW* (2008).

[37] P. Kotsampopoulos, A. Rigas, J. Kirchhof, G. Messinis, A. Dimeas, N. Hatziargyriou, V. Rogakos and K. Andreadis, "EMC issues in the interaction between smart meters and power electronic interfaces," *IEEE Trans. Power Del.*, article in press.

[38] Q. Sun, H. Li, Z. Ma, C. Wang, J. Campillo, Q. Zhang, F. Wallin and J. Guo, "A Comprehensive Review of Smart Energy Meters in Intelligent Energy Networks," *IEEE Internet Things J.*, article in press.

**Panayiotis Moutis** (S'07, M'16) received the Diploma and Phd in Electrical and Computer Engineering (ECE) from the National Technical University of Athens (NTUA), Greece, in November 2007 and January 2015, respectively. Since 2006 he has been offering technical consulting to the sectors of photovoltaic investments and sustainability in Greece from various posts. In 2014 he served as a Research Fellow on Microgrids at the University of Greenwich, in collaboration with Arup. Since February 2016 he is employed as a Postdoctoral Research Associate at Carnegie Mellon University. His research interests lie in the field of renewable sources integration, virtual power plants, microgrids and application of artificial intelligence to power system management and control. He is member of the Power & Energy, Industrial Electronics, Computational Intelligence and Computer IEEE societies and a registered member of the Technical Chamber of Greece.

**Pavlos S. Georgilakis** (M'01, SM'11) received the Diploma in Electrical and Computer Engineering and the Ph.D. degree from the National Technical University of Athens (NTUA), Athens, Greece in 1990 and 2000, respectively. He is currently Assistant Professor at the School of Electrical and Computer Engineering of NTUA. His research interests include power systems optimization, smart grids, and distributed energy resources.

**Nikos D. Hatziargyriou** (S'80–M'82–SM'90–F'09) is Professor with the School of Electrical and Computer Engineering at the National Technical University of Athens (NTUA), and CEO of the Hellenic Electricity Distribution Network Operator (HEDNO). From 2007 to 2012, he was Deputy CEO of the Public Power Corporation (PPC) in Greece, responsible for Transmission and Distribution Networks, island DNO, and the Center of Testing, Research and Prototyping. His research interests include smart grids, distributed energy resources, microgrids, renewable energy sources and power system security. Prof. Hatziargyriou is the Editor-in-Chief of the IEEE Transactions on Power Systems, a Consulting Editor of the IEEE Transactions on Sustainable Energy, past-Chair of the IEEE Power System Dynamic Performance Committee, member of CIGRE, Chair of Study Committee C6 of CIGRE, member of the Board of Directors of EURELECTRIC, and Chair of the EU Technology Platform on SmartGrids.